\documentclass[natbib2009,preprint2]{aastex6}

\usepackage{natbib}
\usepackage{amsmath}
\usepackage{multirow}
\usepackage{color}

\newcommand{\hi}{\mbox{H{\small I}}}
\newcommand{\hii}{\ion{H}{2}}
\newcommand{\cii}{\mbox{[\ion{C}{2}]}}
\newcommand{\ci}{\mbox{\ion{C}{1}}}

\newcommand{\nit}{[\ion{N}{2}]}

\newcommand{\niii}{[\ion{N}{2}]~205~$\mu$m}

\newcommand{\pt}{$P_{\rm th}$}
\newcommand{\ptot}{$P_{\rm tot}$}
\newcommand{\ptotde}{$P_{\rm tot,DE}$}

\newcommand{\pmin}{$P_{\rm min}$}
\newcommand{\pmax}{$P_{\rm max}$}
\newcommand{\ptwo}{$P_{\rm two-phase}$}
\newcommand{\sfrd}{$\Sigma_{\rm SFR}$}
\newcommand{\gasd}{$\Sigma_{\rm gas}$}
\newcommand{\ubar}{$\langle U \rangle$}
\newcommand{\fcnm}{$f_{\rm CNM}$}
\newcommand{\fion}{$f_{\rm ion}$}
\newcommand{\fhd}{$f_{\rm H_2,diff}$}

\shorttitle{Thermal Pressure in the Cold Neutral Medium of Nearby Galaxies}
\shortauthors{Herrera-Camus et al.}

\begin{document}

\title{Thermal Pressure in the Cold Neutral Medium of Nearby Galaxies}


\author{R.~Herrera-Camus\altaffilmark{1,2}, A.~Bolatto\altaffilmark{2}, M.~Wolfire\altaffilmark{2}, E.~Ostriker\altaffilmark{3}, B.~Draine\altaffilmark{3}, A.~Leroy\altaffilmark{4}, K.~Sandstrom\altaffilmark{5}, L.~Hunt\altaffilmark{5}, R.~Kennicutt\altaffilmark{7}, D.~Calzetti\altaffilmark{8}, J.D.~Smith\altaffilmark{9}, K.~Croxall\altaffilmark{4}, M. Galametz\altaffilmark{10}, I. de Looze\altaffilmark{11}, D. Dale\altaffilmark{12}, A. Crocker\altaffilmark{13}}
\author{B. Groves\altaffilmark{14}}

\altaffiltext{1}{Max-Planck-Institut f\"{u}r Extraterrestrische Physik (MPE), Giessenbachstr., D-85748 Garching, Germany} 
\altaffiltext{2}{Department of Astronomy, University of Maryland, College Park, MD 20742, USA}
\altaffiltext{3}{Department of Astrophysical Sciences, Princeton University, Princeton, NJ 08544, USA}
\altaffiltext{4}{Department of Astronomy, The Ohio State University, 4051 McPherson Laboratory, 140 West 18th Avenue, Columbus, OH 43210, USA}
\altaffiltext{5}{Center for Astrophysics and Space Sciences, Dept. of Physics, Univ. of California, San Diego, 9500 Gilman Dr, La Jolla, CA 92093, USA}
\altaffiltext{6}{INAF-Osservatorio Astrofisico di Arcetri, Largo E. Fermi 5, I-50125 Firenze, Italy}
\altaffiltext{7}{Institute of Astronomy, University of Cambridge, Madingley Road, Cambridge CB3 0HA, UK}
\altaffiltext{8}{Department of Astronomy, University of Massachusetts, Amherst, MA 01003, USA}
\altaffiltext{9}{Department of Physics and Astronomy, University of Toledo, 2801 West Bancroft Street, Toledo, OH 43606, USA}
\altaffiltext{10}{European Southern Observatory, Karl Schwarzschild Strasse 2, D-85748 Garching, Germany}
\altaffiltext{11}{Department of Physics and Astronomy, University College London, 132 Hampstead Road, London, NW1 2PS, UK}
\altaffiltext{12}{Department of Physics and Astronomy, University of Wyoming, Laramie, WY 82071, USA}
\altaffiltext{13}{Department of Physics, Reed College, Portland, OR 97202, USA}
\altaffiltext{14}{Research School of Astronomy \& Astrophysics, Australian National University, Canberra, ACT 2611, Australia}

\begin{abstract}
Dynamic and thermal processes regulate the structure of the multi-phase interstellar medium (ISM), and ultimately establish how galaxies evolve through star formation. Thus, to constrain ISM models and better understand the interplay of these processes, it is of great interest to measure the thermal pressure (\pt) of the diffuse, neutral gas. By combining  \cii~158~$\mu$m, \hi, and CO data from 31 galaxies selected from the {\it Herschel} KINGFISH sample, we have measured thermal pressures in 534 predominantly atomic regions with typical sizes of $\sim$1~kiloparsec. We find a distribution of thermal pressures in the $P_{\rm th}/k\sim10^{3}-10^{5}$~K~cm$^{-3}$ range. For a sub-sample of regions with conditions similar to those of the diffuse, neutral gas in the Galactic plane, we find thermal pressures that follow a log-normal distribution with a median value of $P_{\rm th}/k\approx3600$~K~cm$^{-3}$. These results are consistent with thermal pressure measurements using other observational methods. We find that \pt\ increases with radiation field strength and star formation activity, as expected from the close link between the heating of the gas and the star formation rate. Our thermal pressure measurements fall in the regime where a two-phase ISM with cold and warm neutral medium could exist in pressure equilibrium. Finally, we find that the midplane thermal pressure of the diffuse gas is about $\sim30\%$ of the vertical weight of the overlying ISM, consistent with results from hydrodynamical simulations of self-regulated star formation in galactic disks. 
\end{abstract}


\section{{\bf Introduction}} 


In the widely accepted thermal instability model of the multiphase interstellar medium (ISM), most of the neutral atomic gas resides in two distinct phases commonly referred as the cold neutral medium (CNM; $T\lesssim300$~K), and the warm neutral medium (WNM; peak temperature around $T\approx8000$~K) \citep{rhc_field69,rhc_cox05,rhc_heiles03}. These two phases coexist in pressure equilibrium in a relatively narrow range of pressure \citep[$P_{\rm min} < P < P_{\rm max}\approx3P_{\rm min}$;][]{rhc_field69} regulated by the thermal balance between heating and radiative cooling \citep{rhc_draine78,rhc_wolfire95,rhc_wolfire03}, and the vertical pressure exerted by the gravitational field \citep{rhc_badhwar77,rhc_ostriker10,rhc_kim11}. 

The characteristics of the thermal pressure (\pt) curve depend directly on the gas heating rate, which in turn is a function of the metallicity, the ionization rate of atomic hydrogen and the FUV radiation field \citep[e.g.,][]{rhc_wolfire95,rhc_wolfire03}. The latter is directly proportional to the star formation activity and illustrates the strong connection between pressure and star formation in the ISM. \cite{rhc_blitz06} find that the hydrostatic (or dynamical equilibrium) pressure is empirically correlated with the fraction of the neutral gas that is molecular and therefore available to form stars. \cite{rhc_ostriker10} and \cite{rhc_kim11,rhc_kim13} consider the connection between thermal pressure and star formation on $\sim$kpc scales for a model in which the disk evolves into a state of dynamical, thermal, and star formation equilibrium. The analytic model of \cite{rhc_ostriker10} hypothesizes that if the midplane thermal pressure is higher than $P_{\rm max}$ or lower than $P_{\rm min}$, the fraction of the cold gas and the star formation activity evolve in order to bring the midplane pressure close to the average value defined by the two-phase thermal pressure curve. The numerical hydrodynamic ISM/star formation simulations of \cite{rhc_kim11,rhc_kim13} support that these hypotheses are satisfied for a range of galactic environments.

In the Galactic plane, different observational techniques have been employed to characterize the distribution of thermal pressures of the diffuse, neutral gas. \cite{rhc_jenkins01,rhc_jenkins11} use ultraviolet spectra of local stars to identify absorption features created by neutral carbon (\ci). These features can be used to determine the population ratios between the three fine-structure ground electronic state levels of \ci, revealing the excitation conditions and thermal pressure of the diffuse gas along the line of sight. \cite{rhc_jenkins11} find a distribution of thermal pressures that can be well represented by a log-normal distribution that extends from $P_{\rm th}/k\sim10^{2}$ to $10^{4.5}$~K~cm$^{-3}$, with a mean value of $P_{\rm th}/k\approx3800$~K~cm$^{-3}$. \cite{rhc_goldsmith13}, based on ultraviolet measurements of interstellar CO towards nearby stars \citep{rhc_sheffer08}, calculate thermal pressure values for diffuse interstellar molecular clouds in the $4600-6800$~K~cm$^{-3}$ range. An additional method to probe the diffuse gas along a particular line of sight is to use \cii~158~$\mu$m velocity-resolved observations towards bright infrared continuum sources. This allows, based on the absorption and emission features in the spectra, a measure of the line opacity and the line peak temperature, which in turn can be used to derive the density and thermal pressure of the neutral gas. Using this technique, \cite{rhc_gerin15} find a median thermal pressure of $P_{\rm th}/k\approx5900$~K~cm$^{-3}$ in 13 lines of sight in the Galactic plane.

On the modeling side, \cite{rhc_wolfire03} use a comprehensive approach that considers the different sources of heating and cooling of the gas in order to estimate $P_{\rm min}$, $P_{\rm max}$ and the average thermal pressure in the Galactic plane as a function of radius. They conclude that most of the neutral gas in the ISM of the Galaxy out to $\sim18$~kpc have thermal pressures that lie between $P_{\rm min}$ and $P_{\rm max}$ \citep[standard $P_{\rm min}$ and $P_{\rm max}$ values in the Galactic plane are approximately $2\times10^3$ and $5\times10^3$~K~cm$^{-3}$, respectively;][]{rhc_wolfire03}. Inside the solar circle, they calculate a mean thermal pressure of $P_{\rm th}/k\approx3000$~K~cm$^{-3}$, which is lower than, but consistent with, observational results \citep{rhc_jenkins11,rhc_gerin15}.

In this paper we use a method developed by \cite{rhc_kulkarni87} that combines the \cii~158~$\mu$m and \hi~21~cm line to measure the \cii-cooling rate, the density of the neutral gas and, for a given temperature of the CNM, the thermal pressure in the neutral ISM of nearby galaxies. For CNM temperatures in the range $\sim40-400$~K, the results are quite insensitive to the adopted temperature (e.g., see Figure~\ref{P_curves}). This paper is organized as follows. In Section~2 we describe the sample of galaxies and the data. In Section~3 we discuss our method to measure thermal pressures using the \cii\ and \hi\ data. In Section~4 we describe our region selection criteria and the assumptions made in the thermal pressure calculation. In Section~5 we analyze the resulting distribution of thermal pressures. In Section~6 we explore the connection between thermal pressure and star formation activity (and radiation field strength). In Section 7 we measure the total to thermal pressure ratio and we compare it to theoretical predictions. Finally, in Section~8 we present our summary and conclusions.

\section{{\bf Main Sample Description}} 

Our sample consists of 31 galaxies drawn from the KINGFISH sample \citep[``Key Insights on Nearby Galaxies: A Far-Infrared Survey with Herschel'';][]{rhc_kennicutt11} that have CO and \hi\ observations available from the THINGS \citep[``The HI Nearby Galaxy Survey'';][]{rhc_walter08} and HERACLES \citep[``The HERA CO Line Extragalactic Survey'';][]{rhc_leroy09} surveys. See Table~B1 in the Appendix for a list of the galaxies. With the exception of NGC~3077, that is classified as I0~pec, all the other galaxies in our sample are spirals. They span a range in total infrared (TIR) luminosity of $L_{\rm TIR}\sim10^{8.3}-10^{10.7}$~$L_{\odot}$ \citep{rhc_dale12} and in distance of $D\sim2.8-26.5$~Mpc. Their metallicities, taken from \cite{rhc_moustakas10} and measured as the average between the characteristic oxygen abundances from the \cite{rhc_pilyugin05} (PT05) and \cite{rhc_kobulnicky04} (KK04) calibrations \citep{rhc_croxall13}, are in the $12 + \text{log(O/H)} \sim 8.1 - 9.0$ range. 

\subsection{KINGFISH \cii}

In this work we use \cii~158~$\mu$m observations drawn from the {\it Herschel} key program  KINGFISH \citep{rhc_kennicutt11}. These were carried out with the Photodetector Array Camera \& Spectrometer (PACS) on board {\it Herschel}, and were reduced using the Herschel Interactive Processing Environment (HIPE) version 11.0. For more details on the data reduction process we refer to \cite{rhc_croxall13}. The angular resolution of the PACS spectrometer at 158~$\mu$m is $\approx12\arcsec$. More than half of the \cii\ maps consist of a strip that covers the central region of the galaxy and part of the disk. In addition, there are cases where we have coverage of extra-nuclear regions located in the outskirts of the disk (e.g. M~101, NGC~6946). 

\subsection{THINGS \hi}

We retrieve \hi\ maps from the Very Large Array THINGS survey \citep{rhc_walter08} and a collection of new and archival Karl G. Jansky Very Large Array (VLA) data \citep{rhc_schruba11,rhc_leroy13}. These have angular resolutions in the $6\arcsec-25\arcsec$ range. For more details on the data reduction and map properties we refer to \cite{rhc_walter08}. We convert the 21~cm intensities into \hi\ surface densities via

\begin{equation}
\Sigma_{\rm HI}~[M_{\odot}~{\rm pc}^{-2}] = 0.02~I_{\rm HI}\times{\rm cos}~i~[{\rm K~km~s^{-1}}].
\end{equation}

\noindent This equation assumes optically thin emission, includes a factor 1.36 to account for the contribution from Helium and is projected to face-on orientation using inclinations ($i$) drawn from the compiled list in \cite{rhc_kennicutt11} and \cite{rhc_hunt15}. 

\subsection{HERACLES CO}

We trace the molecular gas using CO($J=2\rightarrow1$) observations taken with the Heterodyne Receiver Array (HERA) on the IRAM 30~m telescope obtained as part of the HERACLES survey \citep{rhc_leroy09}. The angular resolution of the HERACLES data is about $\sim13$\arcsec, similar to the resolution of {\it Herschel} using the PACS \cii\ observations. We convert the CO($2\rightarrow1$) intensities into molecular mass surface densities following

\begin{equation} \label{Sigma_mol}
\Sigma_{\rm mol}~[M_{\odot}~{\rm pc}^{-2}] = 6.25~I_{\rm CO}\times{\rm cos}~i~[{\rm K~km~s^{-1}}],
\end{equation}

\noindent where we have assumed a CO line ratio of $I_{\rm CO}(2\rightarrow1)/I_{\rm CO}(1\rightarrow0)=0.7$ \citep{rhc_leroy12} and a standard Milky~Way conversion factor $\alpha_{\rm CO} = 4.4$~M$_{\odot}$~pc$^{-2}$~(K~km~s$^{-1}$)$^{-1}$ equivalent to $X_{\rm CO}=2.0\times10^{20}$~cm$^{-2}$~(K~km~s$^{-1}$)$^{-1}$. This assumption may not be correct for the central kiloparsec region of KINGFISH galaxies where $\alpha_{\rm CO}$ tends to be a factor $\sim2$ lower than the galaxy mean \citep{rhc_sandstrom13}. This should not be a problem for this study: our analysis only focus on predominantly atomic regions (see Section~4), therefore we exclude the molecular-dominated central kiloparsec regions of KINGFISH galaxies. Note that Equation~(\ref{Sigma_mol}), as for \hi, includes a factor of 1.36 to account for helium and $cos~i$ to correct for inclination. HERACLES is sensitive to molecular gas mass surface densities down to a $3\sigma$ level of $\Sigma_{\rm mol}\sim4$~M$_{\odot}~{\rm pc}^{-2}$.

\subsection{Additional Data}

In order to trace the obscured and un-obscured components of the star formation activity in our galaxies we use a combination of the 24~$\mu$m and H$\alpha$ emission. The 24~$\mu$m maps were drawn from the Spitzer Infrared Nearby Galaxy Survey \citep[SINGS;][]{rhc_kennicutt03}. The H$\alpha$ images were assembled and processed by \citep{rhc_leroy12} and come mainly from the SINGS and Local Volume Legacy \citep{rhc_dale09} surveys, but are also retrieved from GOLDMine \citep{rhc_gavazzi03} and the Palomar Las Campanas Atlas \citep{rhc_boselli02,rhc_knapen04,rhc_hoopes01}. The H$\alpha$ data were corrected for Galactic extinction \citep{rhc_schlegel98}, foreground stars were masked \citep{rhc_mmateos09b} and the \nit\ contribution was removed \citep{rhc_kennicutt08b,rhc_kennicutt09}. We also use {\it Spitzer}/IRAC SINGS 3.6~$\mu$m maps to measure stellar mass surface densities (see Appendix). Finally, we use {\it Herschel} PACS 70 and 160~$\mu$m maps drawn from the photometric KINGFISH sample \citep{rhc_dale12}.

\begin{figure*}
\epsscale{1.1} 
\plotone{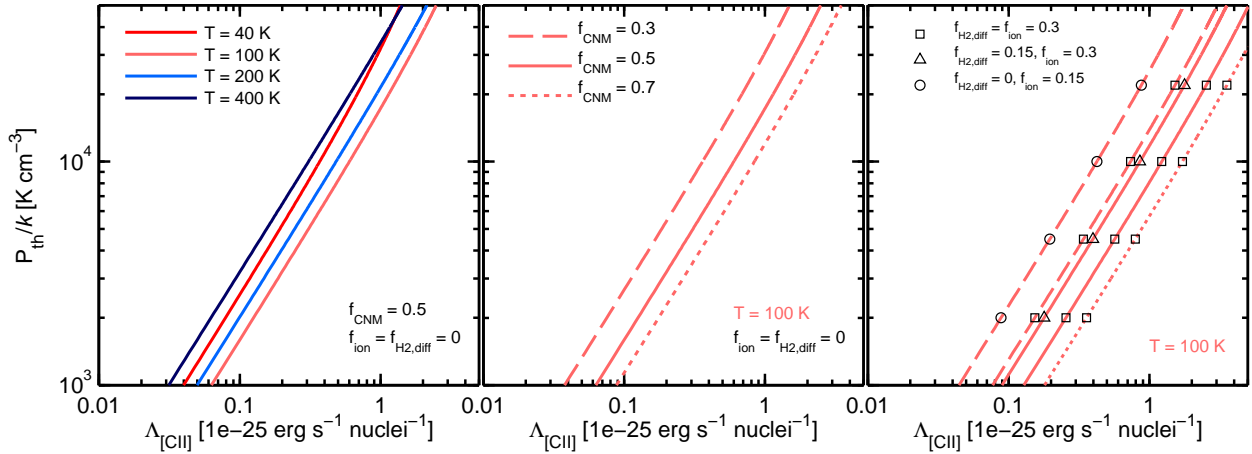}
\caption{Thermal pressure $P_{\rm th}$ as a function of the \cii\ cooling rate per H nucleon ($\Lambda_{\rm [CII]}$; Equation~\ref{eq: n_CII}) for different assumptions on the temperature of the gas ($T$), the fraction of the total $N_{\rm H}$ associated to the CNM phase ($f_{\rm CNM}$), the fraction of the observed \cii\ intensity arising from the ionized gas ($f_{\rm ion}$), and the fraction of the diffuse ISM mass contributed by diffuse, ``CO-dark'' H$_{2}$ ($f_{\rm H_2,diff}$). For all curves we have assumed gas-phase carbon abundance of ${\rm (C/H)_{\rm gas}=1.5\times10^{-4}}$ \citep{rhc_gerin15} and an electron fraction of $n_{\rm e}/n_{\rm H}=4.3\times10^{-4}$ \citep{rhc_draine_book}. {\it (Left Panel)} For a fixed $f_{\rm CNM}=0.5$ and $f_{\rm ion}=f_{\rm H_2,diff}=0$ we vary the temperature of the gas for four different values: $T=40$~K (red), $T=100$~K (orange), $T=200$~K (blue), and  $T=400$~K (dark blue). {\it (Middle Panel)} For a fixed temperature of $T=100$~K and $f_{\rm ion}=f_{\rm H_2,diff}=0$ we vary the CNM fraction: $f_{\rm CNM}=0.3$ (dashed), $f_{\rm CNM}=0.5$ (solid), and $f_{\rm CNM}=0.7$ (dotted). {\it (Right Panel)} For a fixed gas temperature of $T=100$~K, we show cases where we vary $f_{\rm CNM}$ (keeping the same line convention that in the middle panel), and $f_{\rm ion}$ and $f_{\rm H_2,diff}$ by assuming the following cases: $f_{\rm ion}=f_{\rm H_2,diff}=0.3$ (square), $f_{\rm ion}=0.3, f_{\rm H_2,diff}=0.15$ (triangle), and $f_{\rm ion}=0.15, f_{\rm H_2,diff}=0$ (circle).\label{P_curves}}
\end{figure*}

\subsection{Methods}

\subsubsection{Convolution of the data to a common resolution}

We convolve all of our maps to have the same angular resolution. The final angular resolution varies from galaxy to galaxy, as the \hi\ map beam size can be higher or lower than the angular resolution of the \cii, CO and PACS~160~$\mu$m maps ($\approx13\arcsec$). The H$\alpha$, 24~$\mu$m and PACS 70~$\mu$m maps all have higher resolution than the \cii\ and CO data. The common angular resolutions to which all maps of a particular galaxy were convolved are listed in Table~B1. After the convolution, the physical sizes of our regions have a median size of $\approx0.8$~kpc.

\subsubsection{Star formation rate measurements}

We measure star formation rate surface densities ($\Sigma_{\rm SFR}$) using a combination of the H$\alpha$ and 24~$\mu$m emission following the calibration by \cite{rhc_calzetti07} (Equation~8), which is optimized for resolved regions rather than global galaxies \citep{rhc_kennicutt09}. This calibration adopts a truncated Salpeter IMF with slope 1.3 in the range $0.1-0.5$~M$_\odot$ and slope 2.3 in the range $0.5-120$~M$_\odot$.

\subsubsection{Dust-weighted mean starlight intensity, \ubar, from the Draine \& Li dust modeling} 

The thermal pressure of the diffuse, neutral gas is proportional to the FUV radiation field strength \citep{rhc_wolfire03}. As we discuss in Section~6, the latter can be connected to the dust-weighted mean starlight intensity as calculated in the \cite{rhc_draine07} model.
In this model dust is exposed to a range of radiation fields that give rise to two components: (1) the ``Photodissociation region (PDR) component'' where a fraction $\gamma$ of the dust mass is heated by a power-law distribution of intensities $U$ over a wide range, $U_{\rm min}\leq U \leq U_{\rm max}$ (and $U_{\rm max} \gg U_{\rm min}$); and (2) a ``diffuse ISM'' component which is heated by a single ($\delta$ function) radiation field, $U = U_{\rm min}$. This component contains most of the dust. The dust-weighted mean starlight intensity is defined as \citep[Equation~17 in][]{rhc_draine07}: 

\begin{equation}
\langle U \rangle = \bigg[(1-\gamma)U_{\rm min}+\frac{\gamma {\rm ln}(U_{\rm max}/U_{\rm min})}{U_{\rm min}^{-1}-U_{\rm max}^{-1}}\bigg].
\end{equation}

We estimate the dust-weighted mean starlight intensity, \ubar, using the 70 to 160~$\mu$m ratio and the empirical fit to the \cite{rhc_draine07} model derived by \cite{rhc_mmateos09}. \ubar\ is normalized to the local interstellar radiation field measured by \cite{rhc_mathis83}. This fit is only valid for regions with $\langle U\rangle\gtrsim0.7$ given that below this value submillimeter data is needed to constrain the dust temperature of the source \citep{rhc_draine07}. 

\section{\bf \cii~158~$\mu$m emission and the thermal pressure in the diffuse, neutral ISM}

The \cii~158~$\mu$m emission is the result of the radiative de-excitation of carbon ions (C$^{+}$) collisionally excited by electrons (e$^{-}$), hydrogen atoms (H) and/or hydrogen molecules (H$_{2}$). Which of these collisional partners dominate the excitation of C$^{+}$ depend on the properties of the gas where \cii\ emission originates. Neutral carbon has a lower ionization potential (11.3~eV) than hydrogen, thus ionized carbon can be found in both neutral (diffuse neutral gas and surface layers of PDRs) and ionized gas phases of the ISM.
The goal of this section is to describe the method by which we can use the observed \cii\ intensity and \hi\ column density to measure the volume density and thermal pressure of the neutral gas in the CNM phase. This requires an identification of the fraction of the observed \cii\ intensity and the column density of H nuclei that is associated with the CNM. In this section we go through this calculation by considering the multi-phase origin of the \cii\ emission, and the contribution to the column density of H nuclei by the CNM, the WNM, and the translucent part of clouds where CO has been dissociated.

In the optically thin limit, the \cii\ integrated line intensity $I_{\rm [CII]}$ resulting from the collisional excitation of C$^{+}$ by a given collisional partner in the neutral or ionized gas is given by \citep[e.g.,][]{rhc_crawford85,rhc_goldsmith12} 

\begin{equation} \label{eq: I_CII}
I_{\rm[CII]} = 2.3 \times10^{-21} \bigg(\frac{2e^{-91.2/T}}{1+2e^{-91.2/T}+(A_{\rm ul}/\sum_{i}R_{{\rm ul},i}n_{i})}\bigg) N_{\rm C^{+}}.
\end{equation}

\noindent Here, $I_{\rm[CII]}$ is in units of erg~s$^{-1}$~cm$^{-2}$~sr$^{-1}$, $T$ is the kinetic temperature in K, $N_{\rm C^{+}}$ is the column density of C$^{+}$ (in cm$^{-2}$) in the C$^+$ region, $A_{\rm ul}$ is the Einstein spontaneous decay rate ($A_{\rm ul}=2.3\times10^{-6}~{\rm s}^{-1}$), $n$ is the volume density of the collisional partner, and $R_{\rm ul}$ is the collisional de-excitation rate coefficient at a kinetic temperature $T$ for a given collisional partner. The sum in the denominator is over collision partners (i.e., H, H$_2$, He or e$^{-}$).  To calculate the value of $R_{\rm ul}$ at a given $T$ we use the expressions in \cite{rhc_goldsmith12} for collisions with hydrogen atoms and electrons, and \cite{rhc_wiesenfeld14} for collisions with hydrogen molecules. 

\renewcommand{\thefootnote}{\fnsymbol{footnote}}

The observed \cii\ intensity, $I_{\rm [CII]}^{\rm obs}$, is the combination of the contributions to the \cii\ emission from the neutral gas ($I_{\rm [CII]}^{\rm neutral}$) and the ionized gas ($I_{\rm [CII]}^{\rm ion}$), thus

\begin{equation} \label{eq: I_CII_comp}
I_{\rm [CII]}^{\rm obs}=I_{\rm [CII]}^{\rm neutral}+I_{\rm [CII]}^{\rm ion}.
\end{equation}

\noindent If we assume that the fraction of the observed \cii\ intensity produced in the ionized gas is $f_{\rm ion}$, then 

\begin{equation} \label{eq: I_CII_comp}
I_{\rm [CII]}^{\rm neutral}=(1-f_{\rm ion})I_{\rm [CII]}^{\rm obs}. 
\end{equation}

For typical CNM and WNM conditions we expect $I_{\rm [CII]}^{\rm CNM}/I_{\rm [CII]}^{\rm WNM}\approx20$\footnote[2]{If we consider a CNM phase with $n_{\rm HI}=50$~cm$^{-3}$ and $T=100$~K, a WNM phase with $n_{\rm HI}=0.5$~cm$^{-3}$ and $T=8000$~K, and comparable column densities in the WNM and CNM, then $R_{ul,\rm HI}({T=\rm100~K})=7.6\times10^{-10}$~cm$^3$~s$^{-1}$, $R_{ul,\rm HI}({T=\rm8000~K})=1.4\times10^{-9}$~cm$^3$~s$^{-1}$ \citep{rhc_goldsmith12} and $I_{\rm [CII]}^{\rm CNM}/I_{\rm [CII]}^{\rm WNM} \approx (50R_{ul,\rm HI}(100)e^{-91.2/100})/(0.5R_{ul,\rm HI}(8000)e^{-91.2/8000})\approx20$.}, so we can neglect the WNM contribution to the total \cii\ emission, i.e., $I_{\rm [CII]}^{\rm neutral}\approx I_{\rm [CII]}^{\rm CNM}.$

\begin{figure*}
\epsscale{1}
\plotone{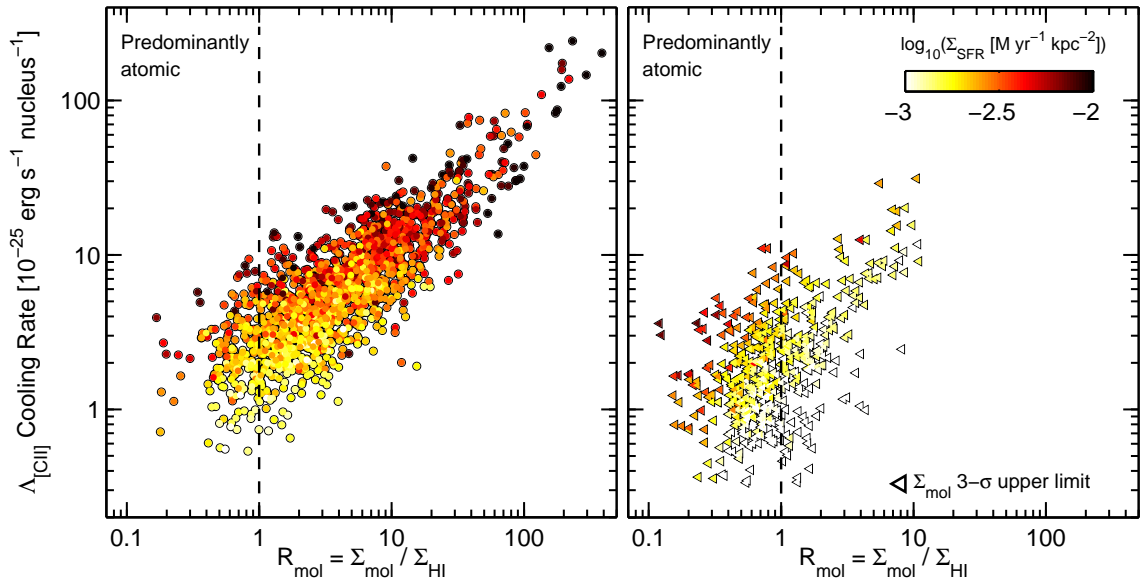}
\caption{\cii\ cooling rate per hydrogen nucleus ($\Lambda_{\rm [CII]}$) as a function of $R_{\rm mol}$, the ratio between the surface density of atomic ($\Sigma_{\rm HI}$) and molecular ($\Sigma_{\rm mol}$) gas. 
The latter employs a CO-based determination of the molecule content, and by definition does not include ``CO-dark'' gas. $\Lambda_{\rm [CII]}$ was calculated using Equation~\ref{eq: n_CII} and assuming $T=100$~K, $f_{\rm CNM}=0.5$, $f_{\rm H_2,diff}=f_{\rm ion}=0$. The left panel shows the regions for which we have detections of [CII], CO and HI emission with $S/N>3$. The right panel, on the other hand, shows the regions for which we only have upper limits driven by non-detections in CO emission. We consider as predominantly atomic regions those with $R_{\rm mol}\leq1$ (dashed line). In these regions, we expect the ${\rm C^{+}}$ collisional excitations to be dominated by collisions with H atoms. The color scale represents the star formation rate surface density ($\Sigma_{\rm SFR}$) of the regions measured as a combination of 24~$\mu$m and H$\alpha$ emission. Regions with lower \cii\ cooling rates tend to show lower $\Sigma_{\rm SFR}$ values. This is expected given that in thermal equilibrium, the heating --proportional to $\Sigma_{\rm SFR}$-- and the cooling --dominated by the \cii\ emission-- are in balance.
\label{atomic_cut}}
\end{figure*}

Now, the \cii\ emission arising from the CNM is the result of collisional excitation by hydrogen atoms, molecules, He and electrons. If we assume a typical ionization fraction for the CNM of $n_{\rm e}/n_{\rm H}=4.3\times10^{-4}$ \citep{rhc_draine_book}, then collisional excitations by electrons can be neglected\footnote[3]{If we assume for the CNM phase an ionization fraction $n_{\rm e}/n_{\rm H}=4.3\times10^{-4}$ \citep{rhc_draine_book} and $R_{ul,\rm e}({\rm 100~K})=1.4\times10^{-9}$~cm$^3$~s$^{-1}$ \citep{rhc_goldsmith12}, then $I_{\rm [CII]}^{\rm HI}/I_{\rm [CII]}^{\rm e} \approx (R_{ul,\rm HI}(100)/R_{ul,\rm e}(100))\times(1/4.3\times10^{-4})\approx7$.}.
The same is true for collisional excitations of C$^+$ by He given that the He collision rate coefficient is only $\sim4$\% of the H collision rate after assuming a cosmic abundance number ratio of ${\rm H/He} = 10$ \citep{rhc_draine_book}.

The CNM consists of the fraction $f_{\rm CNM}$ of the \hi, plus the diffuse, ``CO-dark'' H$_2$. If $N_{\rm H_2,diff}$ is the column density of H$_2$ in the diffuse gas and translucent part of clouds where CO has been dissociated (commonly referred as ``CO-dark'' or ``CO-faint'' gas), then the total column density of H nuclei in the diffuse neutral gas of the ISM is $N_{\rm H}=N_{\rm HI}+2N_{\rm H_2,diff}$. If the \cii\ emission originates primarily from this component, then the \cii\ cooling rate per H nucleon is

\begin{equation} \label{eq: L_CII}
\Lambda_{\rm[CII]}=  \frac{4 \pi I_{\rm[CII]}}{N_{\rm H}}.
\end{equation}

Let $f_{\rm H_2,diff}=2N_{\rm H_2,diff}/(N_{\rm HI}+2N_{\rm H_2,diff})$ be the fraction of the diffuse ISM mass contributed by ``CO-dark'' H$_2$. Then we can express $N_{\rm H}$ as $N_{\rm H}=N_{\rm HI}/(1-f_{\rm H_2,diff})$ and write the \cii\ cooling rate of the diffuse gas in the CNM as

\renewcommand{\thefootnote}{\arabic{footnote}}

\begin{equation} \label{eq: L_CII_v2}
\Lambda_{\rm[CII]}^{\rm CNM} =  \frac{4 \pi I_{\rm[CII]}^{\rm CNM}}{f_{\rm CNM}N_{\rm HI}/(1-f_{\rm H_2,diff})}.
\end{equation}

\noindent Next, combining Equations~(\ref{eq: I_CII}) -- (\ref{eq: L_CII_v2}), and assuming that the atomic and molecular hydrogen C$^{+}$ collisional rates are similar \citep{rhc_goldsmith12,rhc_wiesenfeld14}, we link the observed \cii\ cooling rate per H  nucleon to the CNM neutral gas volume density $(n_{\rm HI}+n_{\rm H_{2}})$ resulting in

\begin{equation}  \label{eq: n_CII}
\begin{split}
\Lambda_{\rm[CII]}^{\rm CNM} & \approx 2.9 \times10^{-20} \bigg(\frac{1-f_{\rm H_2,diff}}{f_{\rm CNM}}\bigg)(1-f_{\rm ion}) \times \\
&  \left(\frac{\rm C}{\rm H}\right)_{\rm gas}\bigg(\frac{2e^{-91.2/T}}{1+2e^{-91.2/T}+(A_{\rm ul}/R_{\rm ul,HI})/(n_{\rm HI}+n_{\rm H_{2}})}\bigg). 
\end{split}
\end{equation}

\noindent The units of $\Lambda_{\rm[CII]}^{\rm CNM}$ are ${\rm erg~s^{-1}~H~nuclei^{-1}}$. Here we have also assumed that all gas-phase carbon is in a singly ionized state, thus $N_{\rm C^{+}}/N_{\rm HI} = {\rm (C/H)_{\rm gas}}$. 

Finally, using Equation~(\ref{eq: n_CII}) we solve for the density of the neutral gas of the CNM, $n_{\rm HI}+n_{\rm H_{2}}$, and then calculate the thermal pressure of the neutral gas following

\begin{equation}  \label{eq: P_th}
P_{th}/k~[{\rm K~cm^{-3}}] = (n_{\rm HI}+n_{\rm H_{2}}+n_{\rm He}) T,
\end{equation}

\noindent where $n_{\rm He}$ is the volume density of helium. For $f_{\rm H_{2}}\lesssim0.5$, and assuming a cosmic abundance number ratio of ${\rm H/He} = 10$, we can approximate $n_{\rm HI}+n_{\rm H_{2}}+n_{\rm He}\approx1.1(n_{\rm HI}+n_{\rm H_{2}})$.

How sensitive is this $\Lambda_{\rm[CII]}$-based thermal pressure measurement to changes in the assumptions of $T$, \fcnm, \fion\ and \fhd? Figure~\ref{P_curves} shows the thermal pressure \pt\ as a function of the cooling rate $\Lambda_{\rm[CII]}$ for different assumptions on $T$ (panel~1), \fcnm\ (panel~2), and \fion\ and \fhd\ (panel~3).  If we only vary the temperature of the CNM gas we find that the resulting thermal pressure varies about a factor of $\sim1.5$ in the $40\gtrsim T \gtrsim100$~K range, and about a factor of $\sim2$ in the $40\gtrsim T \gtrsim400$~K range. We conclude that \cii-based thermal pressure measurements are very robust to even large variations of the CNM temperature. In the case of the CNM fraction, for a given $\Lambda_{\rm[CII]}$ (at fixed $T$) the thermal pressure increases by a factor of $\sim2.3$ if we decrease the CNM fraction from $f_{\rm CNM}=0.7$ to $f_{\rm CNM}=0.3$. This change in the CNM fraction is expected if we go from the inner to the outer parts of the Galactic plane \citep{rhc_pineda13}. Finally, the third panel in Figure~\ref{P_curves} shows that for a given $\Lambda_{\rm[CII]}$ (for fixed assumptions on \fcnm\ and $T$) the resulting thermal pressure decreases by a factor $\sim2$ if we increase the contribution from both, \fion\ and \fhd, from 0 to 30\%.

\section{{\bf Measuring $\Lambda_{\rm[CII]}$ and \pt\ in the KINGFISH sample}} 

Calculating the thermal pressure of the CNM gas based on the \cii\ emission requires selection of regions where the collisional excitation of C$^{+}$ is dominated by the diffuse, neutral gas component. In addition, a number of assumptions need to be made regarding the origin of the \cii\ emission and the temperature and carbon abundance of the CNM gas (see Equation~\ref{eq: n_CII}). In this section we describe our selection criteria for the KINGFISH regions and the assumptions underlying the thermal pressure calculation.

\subsection{Selection of regions}

As a first step to build a sample of quiescent, neutral gas dominated regions we start by identifying those where the mass surface density of atomic gas is higher than that of molecular gas.

Figure~\ref{atomic_cut} shows the \cii\ cooling rate as a function of the molecular ratio, $R_{\rm mol} = \Sigma_{\rm mol}/\Sigma_{\rm HI}$, for 2093 regions for which we have \cii\ and \hi\ detections with $S/N\geq3$. Triangles in the right panel represent regions for which we only have 3-$\sigma$ upper limits on $\Sigma_{\rm mol}$  driven by a non-detection in CO.
In this particular case $\Lambda_{\rm[CII]}$ is calculated using Equation~(\ref{eq: n_CII}) assuming $f_{\rm ion}=f_{\rm H_2,diff}=0$ and $f_{\rm CNM}=0.5$, i.e., half of the atomic gas is in the CNM phase \citep{rhc_heiles03,rhc_pineda13}. The color scale indicates the star formation rate surface density. We find that the \cii\ cooling rate increases as a function of the molecular ratio, and that for a given $R_{\rm mol}$, regions with higher $\Sigma_{\rm SFR}$ have higher \cii\ cooling rates. This is expected in thermal equilibrium, where the heating of the gas powered by the star formation activity is balanced by the cooling, of which the \cii\ emission is one of the main channels \citep[e.g.,][]{rhc_rhc15}.

From a total of 2093 regions with \cii, CO and \hi\ data available, we select for this study of the thermal pressure of the neutral gas 534 regions with ${\rm R_{mol}} \leq 1$ (345 of these regions have upper limits on $\Sigma_{\rm mol}$). Lowering the cut-off level to $R_{\rm mol}\leq0.5$ reduces the number of selected regions to 145, but it does not have a significant effect on the distribution of \cii\ cooling rates.

\begin{figure*}
\epsscale{1.1}
\plotone{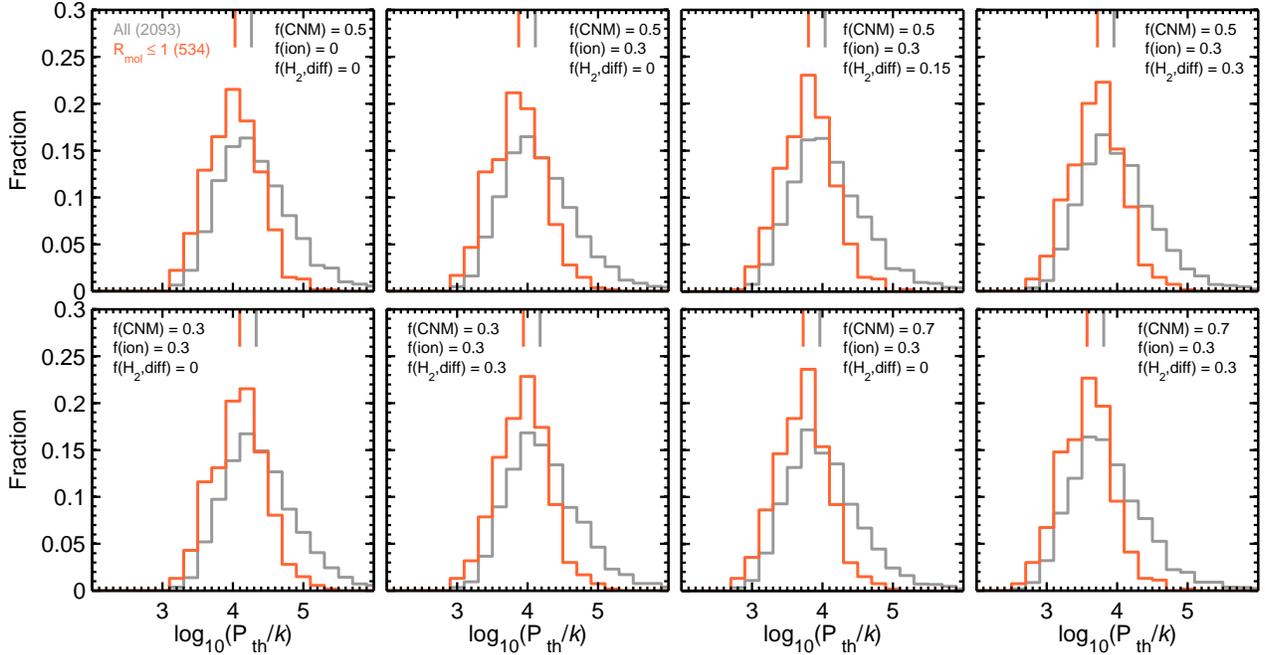}
\caption{Thermal pressure distribution for all regions (grey histograms) and atomic-dominated regions (orange histograms) selected from the KINGFISH sample. Each panel shows the resulting distributions after assuming different values for the fraction of neutral atomic gas in the CNM ($f_{\rm CNM}$), and the contribution to the \cii\ emission from collisional excitations by electrons ($f_{\rm ion}$) and diffuse molecular hydrogen gas ($f_{\rm H_2,diff}$). {\it (Panels 1-4)} We assume $f_{\rm CNM}=0.5$ and $f_{\rm ion}=0.3$ and we increase the value of $f_{\rm H_2,diff}$ from 0 to 0.3. {\it (Panels 5-6)} We assume $f_{\rm CNM}=0.7$ and $f_{\rm ion}=0.3$ and we increase the value of $f_{\rm H_2,diff}$ from 0 to 0.3. For a fixed CNM temperature and carbon abundance, the net effect of increasing $f_{\rm CNM}$, $f_{\rm ion}$ and/or $f_{\rm H_2,diff}$ is to reduce $\Lambda_{\rm[CII]}^{\rm CNM}$ (Equation~\ref{eq: n_CII}). This implies a decrease in the neutral gas density, and as a result, a decrease in the thermal pressure.
}\label{panel_cut_x4}
\end{figure*}

\subsection{Assumptions made in the \pt\ calculation}

As Equation~(\ref{eq: n_CII}) shows, to measure the thermal pressure of the CNM gas we need to make assumptions about the temperature of the CNM gas, the carbon abundance, the fraction of the atomic gas in the CNM, the contribution to the \cii\ emission from the ionized gas, and the mass fraction of diffuse molecular gas mixed with the atomic gas.

\medskip

\noindent {\it CNM temperature:} In this work we assume a CNM temperature of $T=100$~K \citep{rhc_gerin15}. Remember that a change in the temperature assumption of a factor of $\sim2$ around $T=100$~K will only have a small effect ($\lesssim30$\%) on the determination of \pt\ (see Figure~\ref{P_curves}).

\medskip

\noindent {\it Carbon abundance:} To determine the gas-phase carbon abundance of a particular region, we use as a proxy the oxygen abundance of its parent galaxy. Characteristic gas phase oxygen abundances in \hii\ regions of our galaxies were measured by \cite{rhc_moustakas10} based on the theoretical KK04 and empirical PT05 calibration methods. The latter yields metallicities that are systematically lower by about $\sim0.6$~dex compared to those obtained using the KK04 calibration; together, they represent the full range of metallicities one would obtain using other strong-line abundance calibrations \citep{rhc_moustakas10}. For the two calibration methods we convert the O abundances into diffuse depleted carbon abundances using the analytic function in the latest version of the MAPPINGS photoionization code (Nicholls et al. 2016), i.e., 

\begin{equation} \label{eq:Z}
{\rm log(C/H)}\!=\!{\rm log(O/H)}\!+\!{\rm log(10^{-1.00}+10^{(2.72+log(O/H))}}).
\end{equation}

\noindent This analytic function has been re-normalized so that if we input the oxygen gas-phase abundance measured in the Orion nebula \citep[$12+{\rm log(O/H)}=8.65$;][]{rhc_sd11}, we recover a local Galactic depleted ISM carbon abundance of ${\rm (C/H)}=1.5\times10^{-4}$ \citep{rhc_gerin15}. We take as the final carbon abundance the average between the two carbon abundances derived independently from the KK04 and PT05 oxygen metallicities following Equation~(\ref{eq:Z}).

\medskip

\noindent {\it Fraction of the atomic gas in the CNM ($f_{\rm CNM}$)}: Following the results from \cite{rhc_heiles03} and \cite{rhc_pineda13} we assume $f_{\rm CNM}=0.5$ (i.e., half of the atomic gas is in the CNM phase), but we also allow this fraction to vary in the 0.3 to 0.7 range. 

\medskip

\noindent {\it Contribution to the \cii\ emission from ionized gas ($f_{\rm ion}$):} There is a fraction of the observed \cii\ emission that is the result of collisional excitations in \hii\ gas. For a proper calculation of the \pt\ in the neutral gas, this additional contribution needs to be subtracted. One method to account for the contribution from the ionized gas is to use the \niii\ transition. Given that this line arises exclusively from the ionized gas and has a critical density similar to that of the \cii\ line in the ionized medium, the \cii\ to \niii\ ratio is tracer of the fraction of \cii\ emission that originates in the ionized gas, $f_{\rm ion}$. Based on \niii\ observations from the ``Beyond the Peak'' survey --that include multiple regions selected from 21 of the galaxies in our sample-- $f_{\rm ion}$ is estimated to be $f_{\rm ion}\sim0.2-0.4$ for regions with infrared colors $\nu f_{\nu}(70)/\nu f_{\nu}(160)\lesssim1.5$, and $f_{\rm ion}\sim0.1-0.15$ for warmer regions with infrared colors between $1.5\lesssim\nu f_{\nu}(70)/\nu f_{\nu}(160)\lesssim2.5$ (Croxall et al. in prep.). In our sample of predominantly atomic regions 92\% have $\nu f_{\nu}(70)/\nu f_{\nu}(160)\lesssim1.5$, and thus we fix the contribution from the ionized gas to the total \cii\ emission to be $f_{\rm ion}=0.3$. This is consistent with the ionized gas contribution to the total \cii\ luminosity of the Milky Way measured by \cite{rhc_pineda14}.

\medskip 

\noindent {\it Mass fraction of diffuse H$_{2}$ gas mixed with the atomic gas ($f_{\rm H_2,diff}$):} Observations and modeling suggest that 30\% to 50\% of diffuse H$_2$ in the Solar Neighborhood resides in a ``CO-dark'' phase \citep[][]{rhc_wolfire10,rhc_grenier05}. In the Galactic plane \cite{rhc_langer14} find a range of mass fractions of ``CO-dark'' H$_{2}$ in molecular clouds that goes from $\sim20$\% in dense clouds to $\sim75$\% in diffuse molecular clouds. Molecular hydrogen in this ``CO-dark'' gas can contribute to the \cii\ emission by collisionally exciting C$^{+}$ ions with a collisional rate roughly similar to that of hydrogen atoms \citep{rhc_goldsmith12,rhc_wiesenfeld14}. Unfortunately, with our current dataset it is very difficult to constrain the amount of ``CO-dark'' H$_{2}$ gas mixed with the atomic \hi\ gas. The only option we are left with is to correct the \cii\ cooling rate by assuming a mass fraction of diffuse H$_{2}$ gas that is not traced by CO. In this work, and motivated by the observational and modeling results described above, we present results for thermal pressures calculated assuming diffuse H$_{2}$ mass fractions of $f_{\rm H_2,diff}=0$, 0.15, and 0.3.

\begin{figure}
\epsscale{1.1}
\plotone{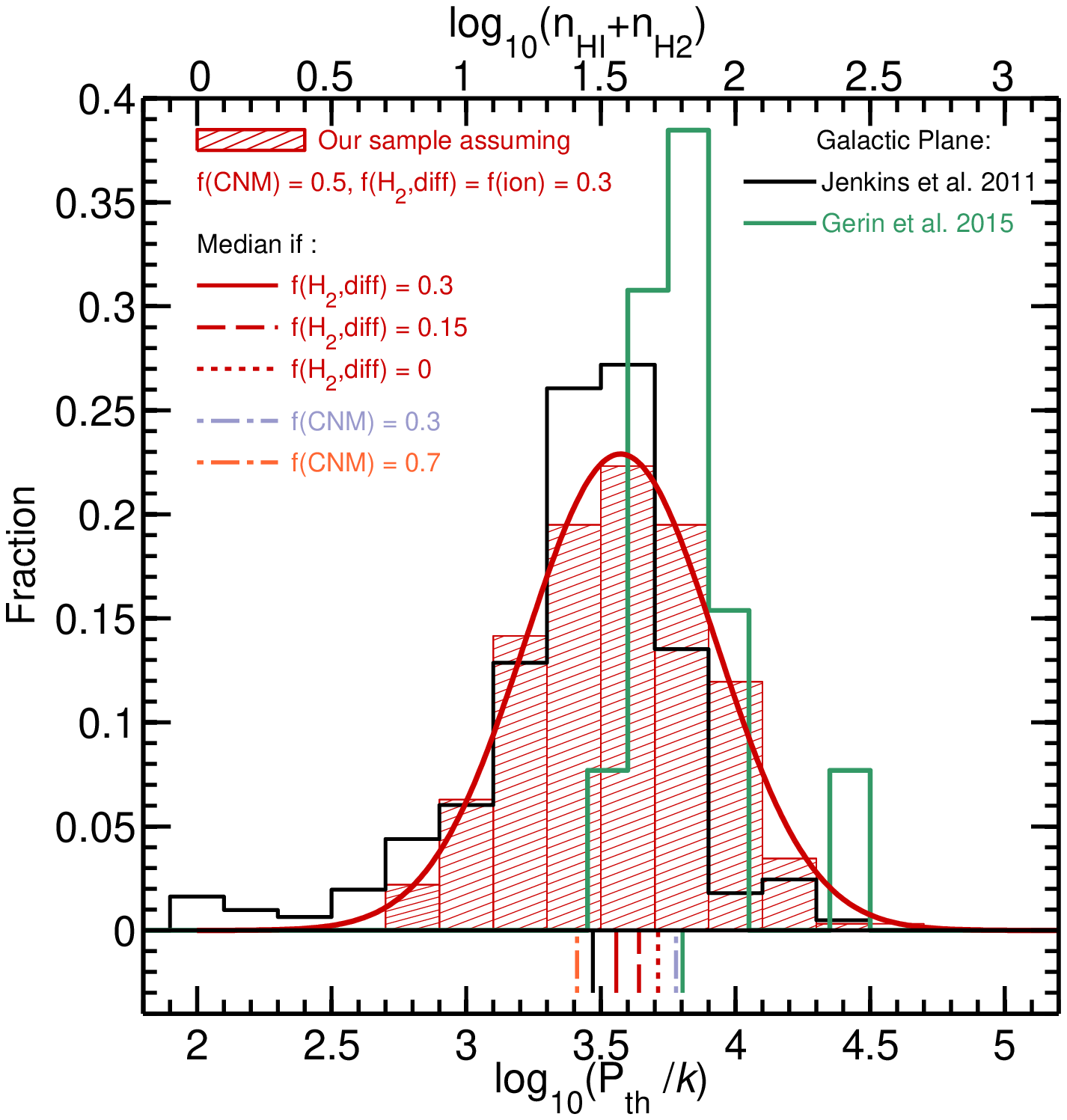}
\caption{Comparison between the thermal distributions from a subsample of KINGFISH regions (318 regions with $\langle U \rangle \leq 3$ and $R_{\rm mol} \leq 1$; red histogram) and regions in the Galactic plane from \cite{rhc_jenkins11} (614 regions, black histogram) and \cite{rhc_gerin15} (13 regions, green histogram). The corresponding value of the diffuse, neutral gas density $n_{\rm HI}+n_{\rm H_2}$ is shown in the upper axis. For the thermal pressure calculations involving the KINGFISH regions we have assumed $f_{\rm CNM}=0.5$, $f_{\rm ion}=0.3$, and $f_{\rm H_2,diff}=0.3$. The KINGFISH distribution of \pt\ can be well represented by the log-normal distribution described in Equation~(\ref{eq: gauss}). In the bottom of the figure we show the median thermal pressure measured in our sample when assuming $f_{\rm CNM}=0.5$, $f_{\rm ion}=0.3$, and $f_{\rm H_2,diff}=0.3$ (red solid line), $f_{\rm H_2,diff}=0.15$ (red dashed line) or $f_{\rm H_2,diff}=0$ (red dotted line). We also include the median values when we vary the CNM fraction, i.e., we assume $f_{\rm ion}=f_{\rm H_2,diff}=0.3$ and $f_{\rm CNM}=0.3$ (purple line) or  $f_{\rm CNM}=0.7$ (orange line). The median thermal pressures measured in \cite{rhc_jenkins11} and \cite{rhc_gerin15} samples are shown in black and green, respectively. \label{P_hist}}
\end{figure}

\section{{\bf Distribution of thermal pressures}}

The thermal pressure distributions resulting from assuming different values of \fcnm, \fion\ and \fhd\ are presented in Figure~\ref{panel_cut_x4}. Each panel shows the results for all regions initially considered for this study (2093 in total), and the 534 regions with molecular ratios $R_{\rm mol}\leq1$. We note that $R_{\rm mol}$ employs a CO-based determination of the molecule content, and by definition does not include ``CO-dark'' gas. In the upper panels of Figure~\ref{panel_cut_x4} we assume that $f_{\rm CNM}=0.5$, and then from left to right we show the resulting thermal pressure distributions if we increase $f_{\rm ion}$ and $f_{\rm H_2,diff}$ contributions from 0 to 30\%. The net effect of increasing $f_{\rm ion}$ and $f_{\rm H_2,diff}$ is that the thermal pressure decreases by a factor of $\sim2$. The bottom panels show the effect on the thermal pressure distributions if we now assume CNM fractions of $f_{\rm CNM}=0.3$ and $f_{\rm CNM}=0.7$ instead of $f_{\rm CNM}=0.5$. We observe that the median \pt\ in the predominately atomic regions when we assume $f_{\rm CNM}=0.3$ is a factor of $\sim2.3$ lower than when we assume $f_{\rm CNM}=0.7$.  The decrease in the thermal pressure with increasing \fion, \fcnm, and/or \fhd\ fractions is expected from Equation~(\ref{eq: n_CII}). The [CII] cooling rate per H nucleon of the diffuse CNM ($\Lambda_{\rm[CII]}^{\rm CNM}$) is proportional to $I_{\rm [CII]}^{\rm CNM}/N_{\rm H}$. Therefore, reducing $I_{\rm [CII]}^{\rm CNM}$ by increasing \fion, or increasing $N_{\rm H}$ by increasing \fcnm\ and/or \fhd,  will result in a lower value of $\Lambda_{\rm[CII]}^{\rm CNM}$. In the low density limit ($n\ll n_{\rm crit}$), and for a given CNM temperature and carbon abundance, the cooling rate is $\Lambda_{\rm[CII]}^{\rm CNM}\propto(n_{\rm HI}+n_{\rm H_2})\propto P_{\rm th}$, so the thermal pressure increases or decreases at the same rate as the \cii\ cooling rate.

Given that we expect \fion\ to be close to 0.3 (Croxall et al. in prep.), and that individual changes in the assumption of $f_{\rm H_2,diff}$ produce changes that are $\lesssim30$\%, we expect that the comparison of our results to measurements in the Galactic plane and expectations from models can help us to constrain \fcnm. 


\subsection{Comparison to the Galactic distribution of thermal pressures}

We compare the distribution of thermal pressures in our sample of predominantly atomic regions to those observed in the diffuse gas of the Galactic plane \citep{rhc_jenkins11,rhc_goldsmith13, rhc_gerin15}. In the case of \cite{rhc_jenkins11}, they select regions in some measure removed from the influence of bright stars by excluding those where the starlight intensity is $>3$ times the Galactic average. We impose a similar condition based on \ubar\ to the 534 regions with $R_{\rm mol}\leq1$, resulting in a sub-sample of 318 quiescent, predominantly atomic regions. Figure~\ref{P_hist} shows the distribution of thermal pressures for this sub-sample when assuming $f_{\rm CNM}=0.5$, $f_{\rm ion}=0.3$ and $f_{\rm H_2,diff}=0.3$. The distribution can be well represented by a log-normal distribution given by

\begin{equation}\label{eq: gauss}
f(P_{\rm th}/k) = 0.23\times{\rm exp}\bigg(-\frac{(\log_{10}(P_{\rm th}/k)-3.57)^2}{2(0.35)^{2}}\bigg).
\end{equation}

\noindent We measure a median thermal pressure of $P_{\rm th}/k=3610$~K~cm$^{-3}$. Compared to the thermal pressures measured in the Galactic plane, our result is similar to the mean value calculated by \cite{rhc_jenkins11}, and about 50\% lower than the median thermal pressure values calculated by \cite{rhc_goldsmith13} and \cite{rhc_gerin15}. The bottom part of Figure~\ref{P_hist} expands this comparison by including the value of the median thermal pressure in our sample for different assumptions on \fhd\ and \fcnm. We note that we find a better agreement with the \cite{rhc_gerin15} results if we assume $f_{\rm H_2,diff}\leq0.15$ or $f_{\rm CNM}=0.3$. In addition, we can obtain a median thermal pressure value sligthly closer to the one measured by \cite{rhc_jenkins11} if we increase the CNM fraction to $f_{\rm CNM}=0.7$. The latter case is less likely given that CNM fractions greater than $\sim0.5$ are probably too high for diffuse gas. On the other hand, if we assume $f_{\rm CNM}=0.3$ the resulting median thermal pressure is $P_{\rm th}/k=8100$~K~cm$^{-3}$, almost a factor three higher than the median thermal pressure measured by \cite{rhc_jenkins11}. Overall, the comparison between our sample and the Galactic plane results favors ISM properties on $\sim$kiloparsec scales where  $f_{\rm CNM}\sim0.5$ and $0.15\lesssim f_{\rm H_2,diff} \lesssim 0.3$.

\section{{\bf Thermal pressure and star-formation activity}}

\begin{figure*}
\epsscale{1}
\plotone{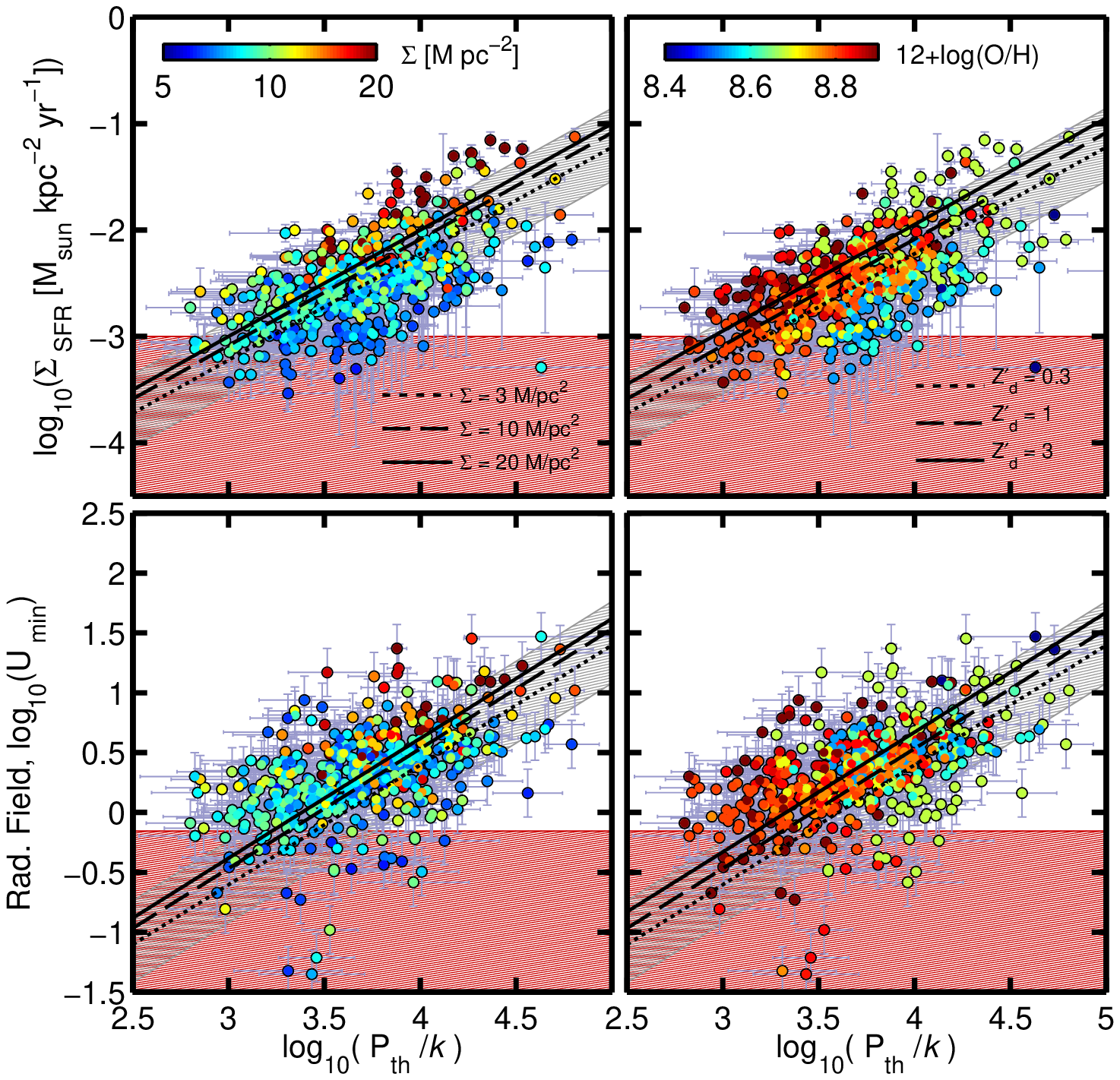}
\caption{Thermal pressure versus star formation surface density (upper panels) and radiation field strength (lower panels) for regions with $R_{\rm mol}\leq1$ in the KINGFISH sample. ({\it Left panels}) The color scale represents the gas surface density $\Sigma_{\rm gas}=\Sigma_{\rm HI}+\Sigma_{\rm H_{2}}$. The results from the \cite{rhc_wolfire03} model (Equation~\ref{eq:2phase_v2} and Equation~\ref{eq:2phase_v3}) when assuming $P_{\rm max}/P_{\rm min}=2$, $Z^{\prime}_{d}=1$ and $\Sigma_{\rm gas}=3$, 10 and 20~M$_{\odot}$~pc$^{-2}$ are shown as dashed, solid and dotted lines, respectively. ({\it Right panels}) Similar to the left panels, but this time the color scale represents the oxygen abundance taken from \cite{rhc_moustakas10}. The model predictions from \cite{rhc_wolfire03} (Equation~\ref{eq:2phase_v2} and Equation~\ref{eq:2phase_v3}) when assuming $P_{\rm max}/P_{\rm min}=2$, $\Sigma_{\rm gas}=10$~M$_{\odot}$~pc$^{-2}$ and $Z^{\prime}_{d}=0.3,1,3$ are shown using dashed, solid and dotted lines, respectively. The hatched grey regions represent the maximal area where $P_{\rm min}\lesssim P_{\rm th}\lesssim P_{\rm max}$ based on the range of values assumed for $P_{\rm max}/P_{\rm min}$, \gasd\ and $Z^{\prime}_{d}$ (the upper edge is defined by $P_{\rm max}/P_{\rm min}=2$, $\Sigma_{\rm gas}=20$~M$_{\odot}$~pc$^{-2}$ and $Z^{\prime}_{d}=3$, and the lower edge by $P_{\rm max}/P_{\rm min}=5$, $\Sigma_{\rm gas}=3$~M$_{\odot}$~pc$^{-2}$ and $Z^{\prime}_{d}=0.3$). The red hatched boxes mark the regions where our star formation rate surface density or radiation field intensity measurements can be affected by additional uncertainties (e.g., in the case of $\Sigma_{\rm SFR}$, contribution from old stars). In all panels we assume that $f_{\rm ion}=f_{\rm H_2,diff}=0.3$. \label{P_sfr}}
\end{figure*}

In the two-phase model for the ISM, the CNM and WNM phases can coexist in pressure equilibrium when the thermal pressure lies within a range set by \pmin\ and \pmax. We define the two-phase pressure as the geometric mean between these two, i.e., $P_{\rm two-phase}=(P_{\rm min}P_{\rm max})^{1/2}$. Hydrodynamical simulations with self-consistent gravitational collapse and star formation feedback to heat and drive turbulence in the ISM find the thermal pressure of the multiphase medium to be close to the \ptwo\ pressure \citep{rhc_kim11,rhc_kim13}. 

Based on the definition of \ptwo\ and using the expression for \pmin\ derived by \cite{rhc_wolfire03}, we can express \ptwo\ as

\begin{multline} \label{eq:2phase_v1}
\frac{P_{\rm two-phase}}{k} \simeq 8500~{\rm K}~{\rm cm}^{-3}~\bigg(\frac{P_{\rm max}}{P_{\rm min}}\bigg)^{1/2} \\
\times \frac{G^{\prime}_{0}Z^{\prime}_{d}/Z^{\prime}_{g}}{1+3.1(G^{\prime}_{0}Z^{\prime}_{d}/\zeta^{\prime}_{t})^{0.365}}.
\end{multline}

\noindent Here, pressure is defined in units of K~cm$^{-3}$, $G^{\prime}_{0}$ corresponds to the mean FUV intensity relative to the value measured locally \citep[$I_{\rm FUV,0}=2.1\times10^{-4}$~erg~cm$^{-2}$~s$^{-1}$~sr$^{-1}$;][]{rhc_draine78}, $Z^{\prime}_{d}$ and $Z^{\prime}_{g}$ are the dust and gas abundances relative to the solar neighborhood values, and $\zeta^{\prime}_{t}$ corresponds to the total cosmic ray/EUV/X-ray ionization rate relative to the value $10^{-16}$~s$^{-1}$. The ratio between \pmin\ and \pmax\ depends on various properties of the gas, including metallicity and the shielding of ionizing radiation. According to \cite{rhc_wolfire95,rhc_wolfire03}, we expect \pmax/\pmin\ to be in the $\sim2-5$ range.

In order to compare our results to the model predictions, we rewrite Equation~(\ref{eq:2phase_v1}) in terms of available observational quantities. Following a similar procedure to that of \cite{rhc_ostriker10}, we first express $G^{\prime}_{0}$ as $G^{\prime}_{0}=I_{\rm FUV}/I_{\rm FUV,0}\approx\Sigma_{\rm SFR}/\Sigma_{\rm SFR,0}$, where $\Sigma_{\rm SFR,0}=2.5\times10^{-3}$~M$_{\odot}$~yr$^{-1}$~kpc$^{-2}$ is the star-formation rate surface density in the solar neighborhood \citep{rhc_fuchs09}. Then, we assume that the total cosmic ray/EUV/X-ray ionization rate $\zeta^{\prime}_{t}$ is proportional to \sfrd\ and inversely proportional to the total gas surface density $\Sigma_{\rm gas}=\Sigma_{\rm HI}+\Sigma_{\rm H_{2}}$. This assumption is discussed in detail in \cite{rhc_wolfire03}, and originates from the fact that cosmic-rays and the hot gas that produces the X-ray emission are related to supernova explosions, while the opacity is related to the surface density of the neutral gas.
Thus, $G^{\prime}_{0}/\zeta^{\prime}_{t}=\Sigma_{\rm gas}/\Sigma_{\rm gas,0}$, where $\Sigma_{\rm gas,0}$ is the surface gas density in the solar neighborhood. For our calculations we assume $\Sigma_{\rm gas,0}=10$~M$_{\odot}$~kpc$^{-2}$ \citep{rhc_wolfire03,rhc_kalberla09}. Finally, we assume that the dust-to-gas ratio follows the metallicity, i.e., $Z^{\prime}_{d}/Z^{\prime}_{g}\approx1$. The new expression for \ptwo\ as function of \sfrd\ is

\begin{multline} \label{eq:2phase_v2}
\frac{P_{\rm two-phase}}{k}\simeq 3.5\times10^{6}~{\rm K}~{\rm cm}^{-3}~\bigg(\frac{P_{\rm max}}{P_{\rm min}}\bigg)^{1/2} \\ 
\times \frac{\Sigma_{\rm SFR}}{\rm M_{\odot}~{\rm yr}^{-1}~{\rm kpc}^{-2}}\times\frac{1}{1+3.1(Z^{\prime}_{d} \Sigma_{\rm gas} /\Sigma_{\rm gas,0})^{0.365}}.
\end{multline}

\noindent The units of \sfrd\ are M$_{\odot}$~yr$^{-1}$~kpc$^{-2}$ and pressure is defined in units of K~cm$^{-3}$.

We derive a third expression for \ptwo, this time expressing $G_{0}$ in terms of the dust-weighted, mean starlight intensity \ubar\ (see Section~2.5.3). The \cite{rhc_mathis83} field integrated between $6-13.6$~eV is related to the Habing field by a factor 1.13, so $G_{0}=1.13U$. In the \cite{rhc_draine07} model the diffuse ISM is exposed to a single radiation field $U=U_{\rm min}$, while a fraction $\gamma$ of the dust mass is heated by a power-law distribution of intensities $U$ over the range $U_{\rm min}\leq U \leq U_{\rm max}$ (where $U_{\rm max}\gg U_{\rm min}$). In our sample of predominately atomic regions $\gamma$ is $\lesssim0.1$ \citep{rhc_draine07b}, which implies that the total power radiated by dust is dominated by dust in the diffuse component of the ISM exposed to $\langle U \rangle \approx U_{\rm min} \approx G_{0}$. Based on this we rewrite 

\begin{multline} \label{eq:2phase_v3}
\frac{P_{\rm two-phase}}{k}\approx 9600~{\rm K}~{\rm cm}^{-3}~\bigg(\frac{P_{\rm max}}{P_{\rm min}}\bigg)^{1/2} \\ 
\times \frac{U_{\rm min}}{1+3.1(Z^{\prime}_{d} \Sigma_{\rm gas} /\Sigma_{\rm gas,0})^{0.365}}.
\end{multline}

In a two-phase ISM model in equilibrium we expect $P_{\rm min} \lesssim P_{\rm th} \lesssim P_{\rm max}$, with \pt\ close to the value of \ptwo. In Figure~\ref{P_sfr} we explore these model expectations by comparing the thermal pressures of our sample of regions with $R_{\rm mol}\leq1$ (assuming $f_{\rm H_2,diff}=0.3$) to the predictions from the \cite{rhc_wolfire03} model. In the left panels we show the correlation between \pt\ and \sfrd\ (upper panel), and  \pt\ and \ubar\ (lower panel). In both cases we use color to indicate the value of \gasd. We include the model results from Equation~(\ref{eq:2phase_v2}) for the $P_{\rm th}-\Sigma_{\rm SFR}$ case, and Equation~(\ref{eq:2phase_v3}) for the $P_{\rm th}-\langle U \rangle$ correlation. In both cases we assume that $P_{\rm max}/P_{\rm min}=2$, $Z^{\prime}_{d}=1$ and three different values for the gas mass surface density, $\Sigma_{\rm gas}=3$ (solid line), 10 (dashed line) and 20~M$_{\odot}$~pc$^{-2}$ (doted). The hatched grey regions represent the maximal area where the condition $P_{\rm min}\lesssim P_{\rm th}\lesssim P_{\rm max}$ is satisfied (this for $2\leq P_{\rm max}/P_{\rm min} \leq 5$, $3 \leq \Sigma_{\rm gas}/({\rm M_{\odot}}~{\rm yr^{-1}}) \leq 20$ and  $0.3 \leq Z^{\prime}_{d} \leq 3$). 
Finally, the red hatched regions mark where $\Sigma_{\rm SFR}\leq10^{-3}$~M$_{\odot}$~yr$^{-1}$~kpc$^{-2}$ or $\langle U \rangle \leq 0.7$, our lower limits for a reliable \sfrd\ or \ubar\ measurements, respectively.

As predicted by the model, we observe a correlation of increasing \pt\ with increasing \ubar\ (Spearman correlation coefficient $\rho=0.53$) and increasing \sfrd\ (Spearman correlation coefficient $\rho=0.60$). This is expected due to the increase of the photoelectric heating as $G_{0}$, which is proportional to $\langle U \rangle$ and $\Sigma_{\rm SFR}$, rises. 

\subsection{Dependence with gas mass surface density \gasd.}

The \cite{rhc_wolfire03} model predicts that for a fixed SFR, the thermal pressure should decrease as a function of \gasd. Recall that $\zeta^{\prime}_{t} \propto (\Sigma_{\rm gas})^{-1}$, so if \gasd\ drops, the electron abundance in the gas rises, which helps to neutralize the charge of the dust grains and thus increase the grain photoelectric heating efficiency. As can be seen from the position of the lines in the left panels of Figure~\ref{P_sfr}, our data are also consistent with this prediction, as we observe that for a fixed \sfrd, regions with higher values of \gasd\ tend to have lower thermal pressures. Finally, it is worth noting that the dispersion in our data ($\sim0.35$~dex) is larger than expected from the model, even if we assume the more extreme cases represented by the grey hatched area. This could be an indication that the gas properties of some regions are still evolving towards dynamical, thermal and star formation equilibrium. As shown in the pressure distributions computed by \cite{rhc_kim11,rhc_kim13} there is always a significant variation about the equilibrium value. On one hand, \sfrd\ varies in time about its mean value, which affects $G_{0}$ and therefore \ptwo. On the other, turbulent compressions and expansions move local regions away from \ptwo.

Another possibility is that part of the observed scatter is driven by observational uncertainties. In particular, \sfrd\ measurements below the $\sim10^{-3}$~M$_{\odot}$~yr$^{-1}$~kpc$^{-2}$ level can be unreliable due to large uncertainties in the H$\alpha$ data and a growing contribution to the 24~$\mu$m emission from old stars \citep[e.g.,][]{rhc_leroy12,rhc_draine14,rhc_rhc15}.

\subsection{Dependence on metallicity}

\renewcommand{\thefootnote}{\fnsymbol{footnote}}

In the right panels of Figure~\ref{P_sfr} we have color coded regions according to the characteristic oxygen abundance of its parent galaxy. For the model comparison, we include results from Equations~(\ref{eq:2phase_v1}) and (\ref{eq:2phase_v2}) assuming $P_{\rm max}/P_{\rm min}=2$, $\Sigma_{\rm gas}=10$~M$_{\odot}$~pc$^{-2}$ and three different values for the dust abundance relative to the solar neighborhood, $Z^{\prime}_{d}=0.3$ (solid line), 1 (dashed line) and 3 (dot-dashed line). The model predicts that for a fixed amount of \sfrd, the thermal pressure increases with decreasing dust abundance $Z^{\prime}_{d}$\footnote[3]{Naively, one might expect lower dust abundance to decrease \pt\ because the photoelectric heating rate decreases.  However, if $Z^{\prime}_{g} \propto Z^{\prime}_{d}$, the cooling rate by fine-structure C and O lines decreases at a compensating rate, as expressed by the numerator of Equation~(\ref{eq:2phase_v1}).  Since in the scenario of lower $Z^{\prime}_{d}$ the X-ray heating rate would remain unchanged, the overall heating/cooling would increase as $Z^{\prime}_{d}$ decreases, raising the equilibrium pressure as expressed by Equation~(\ref{eq:2phase_v2}).}. In our sample, we find that for regions with similar \ubar\ or \sfrd, those with lower metallicities tend to have higher thermal pressures, consistent with the trend predicted in equations~(\ref{eq:2phase_v1}), (\ref{eq:2phase_v2}) and (\ref{eq:2phase_v3}).

\renewcommand{\thefootnote}{\arabic{footnote}}

\section{Comparison between the thermal and total pressure.}

\begin{figure}
\epsscale{1}
\plotone{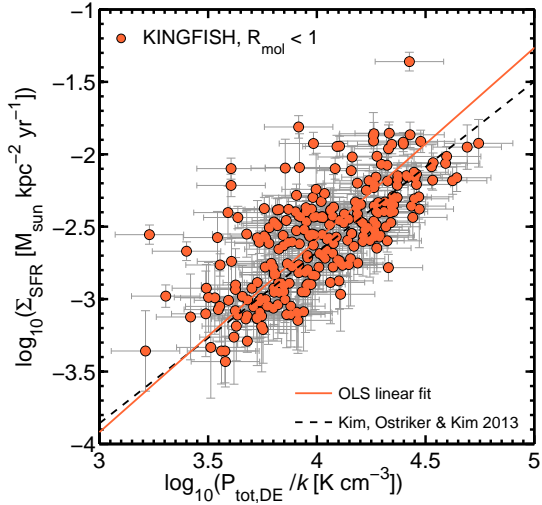}
\caption{Dynamical equilibrium total pressure (\ptotde) versus star formation rate surface density (\sfrd) for regions with $R_{\rm mol}\leq1$ in the KINGFISH sample. The best linear fit to the data, as estimated by the ordinary least-squares (OLS) linear bisector method \citep{rhc_isobe90}, yields ${\rm log}_{10}(\Sigma_{\rm SFR})=1.3\times {\rm log}_{10}(P_{\rm tot,DE}/k)-7.9$ (orange line). We also include the scaling relation between \ptotde\ and \sfrd\ from the hydrodynamical simulations by \cite{rhc_kim13} (black dotted line).
\label{PvsSFR}}
\end{figure}

\begin{figure}
\epsscale{1}
\plotone{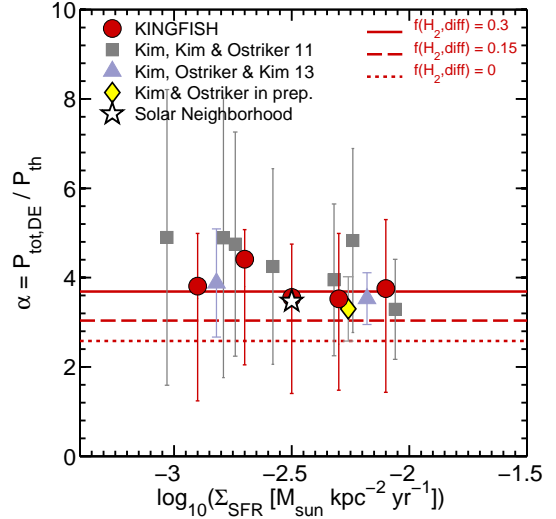}
\caption{Total-to-thermal pressure ratio $\alpha$ as a function of $\Sigma_{\rm SFR}$. The red circles show the median $\alpha$ value measured in our KINGFISH sample of regions with $R_{\rm mol}\leq1$ and assuming $f_{\rm H_2,diff}=0.3$; the horizontal bars represent the 25th to 75th percentile range. The horizontal red line corresponds to the median $\alpha$ value found in our sample across $\Sigma_{\rm SFR}$. If we change the assumption on $f_{\rm H_2,diff}$ to 0 (red dotted line) or 0.15 (red dashed line) we find lower median $\alpha$ values. We also include the results from \cite{rhc_kim11,rhc_kim13} hydrodynamical simulations (grey squares and purple triangles, respectively), Kim \& Ostriker in prep. (yellow diamond), and the solar neighborhood (open star). The markers represent the mean value and the error bars the 1-$\sigma$ standard deviation. 
\label{alpha}}
\end{figure}

In a multi-phase ISM in dynamical equilibrium, the midplane thermal and turbulent pressure forces balance the weight of the ISM, \ptot, that arises from stars, diffuse gas, and dark matter content \citep{rhc_lockman91,rhc_cox05,rhc_ostriker10}. In this context, the ratio between the thermal and total pressure, $\alpha=P_{\rm tot,DE}/P_{\rm th}$, indicates whether the pressure of the diffuse ISM is thermally ($\alpha<2$) or dynamically ($\alpha>2$) dominated. Hydrodynamical simulations of multiphase galactic disks run by \cite{rhc_kim11,rhc_kim13} find an average value of $\alpha\approx4$ (i.e., the thermal pressure is typically $\sim25$\% of the dynamical equilibrium total pressure). These simulations also find a very weak dependence of $\alpha$ with star formation activity \citep[$\alpha \propto (\Sigma_{\rm SFR})^{0.03}$;][]{rhc_kim13}.

Thanks to the wealth of ancillary data available, we combine tracers of thermal pressure, stellar mass surface density, and dark matter content to directly measure $\alpha$. The complete calculation of $\alpha$ is developed in the Appendix. Briefly, we use the expression for the dynamical equilibrium total pressure \ptotde\ detailed in \cite{rhc_kim11}. In this formulation, \ptotde\ is a function of \gasd, the gas vertical velocity dispersion ($\sigma_{z}$), and the midplane density of the stellar disk ($\rho_{*}$) and the dark matter halo ($\rho_{\rm DM}$). Figure~\ref{PvsSFR} shows the resulting \ptotde\ as a function of \sfrd. The correlation agrees well with the best fit to the \cite{rhc_kim13} hydrodynamical simulations (dashed line).  

Regarding the ratio between the total and the thermal pressure, Figure~\ref{alpha} shows the median values of $\alpha$ (and the 25th to 75th percentile range) measured in the KINGFISH regions with $R_{\rm mol}\leq1$ as a function of \sfrd\ (assuming $f_{\rm H_2,diff}=0.3$). Consistent with the expectations from the models, we observe no strong dependence between $\alpha$ and \sfrd\ in the $10^{-3}\lesssim\Sigma_{\rm SFR}/{\rm M}_{\odot}~{\rm yr}^{-1}~{\rm kpc}^{-2}\lesssim10^{-2}$ range, although the dispersion in the data is large. For comparison, we include the approximate $\alpha$ value in the Solar Neighborhood \citep{rhc_wolfire03,rhc_jenkins01} and the mean $\alpha$ values computed in two-dimensional and three-dimensional hydrodynamical simulations by \cite{rhc_kim11} and \cite{rhc_kim13}, respectively.  We have also included results from a new simulation of the solar neighborhood (Kim \& Ostriker in prep.) that follows space-time correlation of supernovae (SNe) with dense and diffuse gas realistically, resolves all thermal phases of the ISM (the hot phase was missing previously), and fully captures the circulation of the galactic fountain. In particular, the amount of radial momentum injection per SN ($p_*$) is self-consistently determined by numerically resolving supernova remnant expansion prior to the onset of cooling \citep{rhc_kim15} in contrast to the previous simulations where it was fixed to $p_*=3\times10^5$~M$_{\odot}$~km~s$^{-1}$.

If we assume $f_{\rm H_2,diff}=0$ we find a mean $\alpha$ of 2.6 (red dotted line), which is about $\sim30$\% lower than the $\alpha$ values resulting from the simulations. In order to find a better agreement between the $\alpha$ values of our regions, the solar neighborhood and the simulations, we need to assume a gas mass fraction of diffuse (``CO-dark'') H$_2$ gas between $f_{\rm H_2,diff}=0.15$ and 0.3. In the latter case we measure a mean $\alpha=3.7$ (red solid line), similar to the mean $\alpha$ found by \cite{rhc_kim13} and Kim \& Ostriker in prep. Independently of our assumption of $f_{\rm H_2,diff}$, we measure a mean $\alpha$ value in our sample that is $\alpha>2$, implying that the ISM is dominated by dynamical processes. 

\section{{\bf Summary and Conclusions}} 

We study the distribution of thermal pressures in the neutral gas of extragalactic regions selected from nearby galaxies that are part of the KINGFISH, HERACLES and THINGS samples. The method we use to measure the thermal pressure relies on the \cii~158~$\mu$m emission that arise from regions where the excitation of C$^{+}$ ions is dominated by collisions with hydrogen atoms  (although it is still important to consider the contribution to the total \cii\ emission from ionized and diffuse, ``CO-dark'' H$_{2}$ gas). In these regions the \cii\ and \hi\ 21~cm line emission can be used to measure the cooling rate per hydrogen nucleus $\Lambda_{\rm [CII]}$ (Equations~\ref{eq: L_CII} and \ref{eq: L_CII_v2}), and then, by assuming a typical temperature for the cold neutral medium and a carbon abundance, invert the cooling equation (Equation~\ref{eq: n_CII}) to infer CNM volume densities; with these we then obtain the thermal pressure \pt\ of the neutral gas (Equation~\ref{eq: P_th}). One advantage of this method is that it is very robust against changes in the assumption of the CNM temperature \citep{rhc_kulkarni87}. 

We highlight the following points: 

\begin{enumerate}
\item We measure the thermal pressure of the neutral gas in 534 predominantly atomic regions ($\sim$1~kpc in size) where we expect the collisional excitation of C$^{+}$ ions to be dominated by the diffuse, neutral gas. These regions were selected from a larger sample of 2093 extragalactic regions with \cii, \hi\ and CO observations available by imposing a cut  $R_{\rm mol}= \Sigma_{\rm mol}/\Sigma_{\rm HI}\leq1$.
The thermal pressure calculations involve a series of assumptions on the properties of the neutral gas and the origin of the \cii\ emission, including the fraction that arises from the ionized gas ($f_{\rm ion}$), and the mass fraction of the total column density that corresponds to H$_{2}$ (``CO-dark'') diffuse gas ($f_{\rm H_2,diff}$). We find a thermal pressure distribution that extends from $P_{\rm th}/k\sim10^3$ to $\sim10^{5}$~K~cm$^{-3}$.

\item We compare the distribution of thermal pressures in our sample to those measured in the neutral, diffuse gas of Galactic plane. For this we follow a similar approach to \cite{rhc_jenkins11} and select a sub-sample of 318 regions with dust-weighted mean starlight intensities $\langle U \rangle \leq 3$. We find that the distribution of thermal pressures in this sub-sample can be well represented by a log-normal distribution. The median thermal pressure is $P_{\rm th}/k\approx3600$~K~cm$^{-3}$, a value that is consistent with those from studies of the diffuse ISM in the Galactic plane by \cite{rhc_jenkins11} and \cite{rhc_gerin15}.
\item The trends observed in the relations between \pt\ and \sfrd\ (and \ubar), as a function of $G_{0}$, \gasd\ and metallicity are consistent with the results from \cite{rhc_wolfire95,rhc_wolfire03} and \cite{rhc_ostriker10} models. In general, the thermal pressures measured in our regions are consistent with the expectations from a two-phase model in pressure equilibrium where \pt\ increases as a function of the radiation field intensity and the star formation activity.
\item We use the thermal pressure together with the midplane gravitational equilibrium pressure of the gas disk to estimate $\alpha=P_{\rm tot,DE}/P_{\rm th}$. We measure mean $\alpha$ values of $\alpha=2.6$ and 3.7 after assuming diffuse H$_{2}$ (``CO-dark'') gas mass fractions of $f_{\rm H_2,diff}=0$ and 0.3, respectively. Irrespective of the assumption on $f_{\rm H_2,diff}$, we find that $\alpha>2$, which implies that the ISM in our regions is dynamically rather than thermally dominated. In order to find optimal agreement between the results from our observations and hydrodynamical simulations by \citep{rhc_kim11,rhc_kim13} it is necessary to assume that the gas mass fraction of diffuse H$_{2}$ (``CO-dark'') gas in our regions is $f_{\rm H_2,diff}\gtrsim0.15$ and the fraction of the neutral gas in the CNM phase is $f_{\rm CNM}\approx0.5$.

\end{enumerate}

We thank the referee for helpful comments and suggestions that improved the paper. We thank C.-G. Kim for helpful discussions and comments. R.H.C. acknowledges support from a Fulbright-CONICYT grant. A.D.B. acknowledges partial support from a CAREER grant NSF-AST0955836, from NSF-AST1139998, from NASA-JPL 1373858, NSF-AST 1412419 and from a Research Corporation for Science Advancement Cottrell Scholar award. M.G.W. was supported in part by NSF grant AST-1411827. E.C.O. is supported by NSF grant AST-1312006. The work of A.K.L. is supported by the National Science Foundation under Grants No. 1615109 and 1615105. PACS has been developed by a consortium of institutes led by MPE (Germany) and including UVIE (Austria); KU Leuven, CSL, IMEC (Belgium); CEA, LAM (France); MPIA (Germany); INAF-IFSI/OAA/OAP/OAT, LENS, SISSA (Italy); IAC (Spain). This development has been supported by the funding agencies BMVIT (Austria), ESA-PRODEX (Belgium), CEA/CNES (France), DLR (Germany), ASI/INAF (Italy), and CICYT/MCYT (Spain). HIPE is a joint development by the Herschel Science Ground Segment Consortium, consisting of ESA, the NASA Herschel Science Center, and the HIFI, PACS, and SPIRE consortia. This work is based (in part) on observations made with Herschel, a European Space Agency Cornerstone Mission with significant participation by NASA. This research has made use of the NASA/IPAC Extragalactic Database (NED), which is operated by the Jet Propulsion Laboratory, California Institute of Technology, under contract with the National Aeronautics and Space Administration. The National Radio Astronomy Observatory is a facility of the National Science Foundation operated under cooperative agreement by Associated Universities, Inc.

\appendix

\section{{\bf Measuring the total to thermal midplane pressure ratio ($\alpha$)}}

In order to measure $\alpha$ we need to take the ratio between the total (\ptot) and the thermal pressure (\pt). The details for the calculation of \pt\ are described in Section~4. For the total pressure \ptot\ we use the formulation described in detail in \cite{rhc_kim11}. For an ISM that is dominated by diffuse gas --and where the effective pressure is dominated by the thermal and turbulent terms  (i.e., cosmic ray, magnetic field, and radiation effects are unimportant)-- the dynamical equilibrium total pressure, \ptotde, can be expressed as \citep[Equation~7 in][]{rhc_kim11}

\begin{equation}
P_{\rm tot,DE} = \frac{\pi G \Sigma^2}{4}\Bigg\{1+\bigg[1+\frac{32\sigma_{z}^2}{\pi^2 G}\frac{\rho_{\rm *,DM}}{\Sigma^2}\bigg]^{1/2} \Bigg\}.
\end{equation}

\noindent Here, $\Sigma$ is the total surface density of the gas, $\sigma_{z}$ is the vertical gas velocity dispersion, and $\rho_{\rm *,DM}$ is the sum of the midplane density of the stellar disk ($\rho_{*}$) and the density of the dark matter halo ($\rho_{\rm DM}$). As shown in \cite{rhc_kim11}, Equation~(A1) can be simplified to

\begin{equation}
P_{\rm tot,DE} = 10^4~k_{B}~{\rm cm}^{-3}~{\rm K} \times \Big(\frac{\Sigma}{10~M_{\odot}~{\rm pc}^{-2}}\Big) \times \Bigg[ 0.33\Big(\frac{\Sigma}{10~M_{\odot}~{\rm pc}^{-2}}\Big) + 1.4\Big(\frac{\rho_{\rm *,DM}}{0.1~M_{\odot}~{\rm pc}^{-3}}\Big)^{1/2}\Big(\frac{\sigma_{z}}{10~{\rm km}~{\rm s^{-1}}}\Big)\Bigg],
\end{equation}

\noindent which is the equation we use to calculate $P_{\rm tot,DE}$ in our sample of regions. Below we describe how we measure the three key parameters --$\rho_{\rm *,DM}$, $\sigma_{z}$ and $\Sigma$-- that go into equation~(A2).

\bigskip

\subsection{Midplane density of the stellar disk ($\rho_{*}$)}

For the calculation of $\rho_{*}$ we follow the procedure described in detail in \cite{rhc_leroy08}. This calculation assumes that the exponential stellar scale height of a galaxy, $h_{*}$, does not vary with radius.

In terms of $h_{*}$ and the stellar mass surface density ($\Sigma_{*}$), the density of the stellar disk $\rho_{*}$ can be expressed as \citep{rhc_vdk88}

\begin{equation}
\rho_{*} = \frac{\Sigma_{*}}{4h_{*}}.
\end{equation}

\noindent We measure $h_{*}$ by assuming that it is related to the stellar scale length, $l_{*}$, by $l_{*}/h_{*}=7.3\pm2.2$ \citep{rhc_kregel02}, and using stellar scale lengths for our galaxies drawn from Table~4 in \cite{rhc_leroy08}. We measure $\Sigma_{*}$ following \cite{rhc_leroy08}

\begin{equation}
\Sigma_{*} = \Upsilon_{*}^{K} \bigg\langle \frac{I_{K}}{I_{3.6}} \bigg\rangle {\rm cos}~i~I_{3.6}, 
\end{equation}

\noindent where $\Upsilon_{*}^{K}$ is the $K$-band mass-to-light ratio and $I_{K}/I_{3.6}$ is the $K$-to-3.6~$\mu$m intensity ratio. Given the large overlap between our sample and the sample of nearby galaxies in \cite{rhc_leroy08}, we use the same values for $\Upsilon_{*}^{K}$ and $I_{K}/I_{3.6}$ adopted by them, i.e., $\Upsilon_{*}^{K}=0.5~M_{\odot}/L_{\odot,K}$ and $I_{K}/I_{3.6}=1.8$. We determine $I_{3.6}$ from {\it Spitzer} 3.6~$\mu$m maps from SINGS \citep{rhc_kennicutt03}. Finally, with estimates of $h_{*}$ and $\Sigma_{*}$, we measure $\rho_{*}$ using Equation~(A3).

\subsection{Dark matter volume density ($\rho_{\rm DM}$)}

We calculate $\rho_{\rm DM}$ by assuming a flat rotation curve for the dark halo (i.e., $V_{c}={\rm constant}$), so $\rho_{\rm DM}$ at a radius $R$ is given by

\begin{equation}
\rho_{\rm DM} = \frac{1}{4\pi G}\bigg(\frac{V_{c}}{R}\bigg)^2.
\end{equation}

\noindent We drew the value of $V_{c}$ for our galaxies from Table~4 in \cite{rhc_leroy08}. These velocities were calculated approximating galaxy rotation curves observed in 21~cm \citep[THINGS,][]{rhc_walter08} following a functional form defined in \cite{rhc_boissier03}. We measure the radial distance of our regions taking into account the inclination of the galaxy \citep{rhc_kennicutt11,rhc_hunt15}.

\subsection{Vertical gas velocity dispersion ($\sigma_{z}$)}

For the vertical gas velocity dispersion we assume a single value of $\sigma_{z}=11$~km~s$^{-1}$ based on the typical gas velocity dispersion value found for the outer, \hi-dominated parts of THINGS galaxies with inclinations lower than $\sim60^{\circ}$ \citep[see Figure~21 in][]{rhc_leroy08}. 

\subsection{Total surface density of the gas ($\Sigma$)}

We measure $\Sigma$ as the sum of the atomic ($\Sigma_{\rm HI}$) and molecular ($\Sigma_{\rm mol}$) gas mass surface densities. See Sections 2.2 and 2.3 for details of the calculation.

\section{{\bf Galaxy Sample}}

\begin{deluxetable}{lccccc}
\tablecaption{List of galaxies included in this study \label{table:sample}}
\tablehead{
\colhead{Source} & \colhead{Distance$^{(a)}$} & \colhead{log$_{10}(M_{*})^{(b)}$} & \colhead{log$_{10}(L_{\rm{TIR}})^{(c)}$} & Resolution$^{(d)}$ & Resolution$^{(e)}$\\
\colhead{} & \colhead{(Mpc)} & \colhead{(M$_{\odot}$)} & \colhead{($L_{\odot}$)} & ($\arcsec$) & (kpc)\\}
\startdata  
IC 2574  & 3.79 & 8.2 & 8.38 & 13.7 & 0.25\\
NGC 337  & 19.3 & 9.32 & 10.07 & 13 & 1.22\\ 
NGC 628  & 7.2 & 9.56 & 9.90 & 20.1 & 0.70\\
NGC 925  & 9.12 & 9.49 & 9.66 & 13 & 0.57\\
NGC 2798  & 25.8 & 10.04 & 10.55 & 28.6 & 1.32\\
NGC 2841  & 14.1 & 10.17 & 10.11 & 11.1 & 0.76\\
NGC 2976  & 3.55 & 8.96 & 8.95 & 13 & 0.22\\
NGC 3049  & 19.2 & 8.58 & 9.54 & 27.8 & 2.59\\
NGC 3077  & 3.83 & 9.34 & 8.8 & 14.3 & 0.26\\
NGC 3184  & 11.7 & 9.5 & 10.04 & 13 & 0.74\\
NGC 3190  & 19.3 & 10.03 & 9.85 & 20.8 & 1.95\\
NGC 3198  & 14.1 & 9.83 & 9.97 & 13 & 0.89\\
NGC 3351  & 9.33 & 10.24 & 9.91 & 13 & 0.59\\
NGC 3521  & 11.2 & 10.69 & 10.54 & 14.1 & 0.77\\
NGC 3627  & 9.38 & 10.49 & 10.45 & 13 & 0.59\\
NGC 3938  & 17.9 & 9.46 & 10.3 & 18.5 & 1.69\\
NGC 4254  & 14.4 & 9.56 & 10.59 & 16.9 & 1.15\\
NGC 4321  & 14.3 & 10.30 & 10.54 & 14.8 & 1.03\\
NGC 4536  & 14.5 & 9.44 & 10.32 & 14.72 & 1.03\\
NGC 4569  & 9.86 & 10.0 & 9.71 & 14.2 & 0.68\\
NGC 4579  & 16.4 & 10.02 & 10.11 & 27.6 & 2.19\\
NGC 4625  & 9.30 & 8.72 & 8.79 & 13 & 0.59\\
NGC 4631  & 7.62 & 9.76 & 10.38 & 14.8 & 0.55\\
NGC 4725  & 11.9 & 10.52 & 9.93 & 18.6 & 1.07\\
NGC 4736  & 4.66 & 10.34 & 9.76 & 13 & 0.29\\
NGC 5055  & 7.94 & 10.55 & 10.34 & 13 & 0.5\\
NGC 5457  & 6.7 & 9.98 & 10.36 & 13 & 0.42\\
NGC 5474  & 6.8 & 8.70 & 8.79 & 20.3 & 0.29\\
NGC 5713  & 21.4 & 10.07 & 10.50 & 15.5 & 1.09\\
NGC 6946  & 6.8 & 9.96 & 10.93 & 13 & 0.36\\
NGC 7331  & 14.5 & 10.56 & 10.72 & 13 & 0.75\\
\enddata
\tablecomments{(a) The method of distance determination is given by \cite{rhc_kennicutt11}; (b) Total infrared luminosity TIR in the $3-1100~\mu$m range. Fluxes are either from \cite{rhc_dale07} or \cite{rhc_dale09}. (c) Stellar masses obtained from the multi-color method described in \cite{rhc_zibetti09}, and listed in \cite{rhc_skibba11}. (d) Common final angular resolution of the convolved \cii, \hi, and CO maps. (e) Linear resolution corresponding to the final angular resolution at the distance of the target.}
\end{deluxetable}

\bibliography{references.bib}

\end{document}